\RequirePackage{fix-cm}
\documentclass[twocolumn]{svjour3}          
\smartqed  

\usepackage{cite}
\usepackage{amsmath,amssymb,amsfonts}
\usepackage{algorithmic}
\usepackage{graphicx}
\usepackage{textcomp}
\usepackage{xcolor}
\usepackage{graphicx}
\usepackage{subcaption}
\usepackage[utf8]{inputenc}

\begin{document}

\title{A pragmatic approach for designing transparent WDM optical networks with multi-objectives 
}


\author{Dao Thanh Hai, Le Anh Ngoc    
}


\institute{F. Author \at
	Post and Telecommunication Institute of Technology \\
	\email{haidt102@gmail.com} 
           \and
           S. Author \at
           Electric Power University \\ 
           \email{anhngoclevn@gmail.com}    
}

\date{Received: date / Accepted: date}

\maketitle

\begin{abstract}
The era of Artificial Intelligence and Internet of Things are underway, pushing the proliferation of bandwidth-intensive services such as virtual realities, big data exchanges and autonomous vehicles.  In this context, optical transport networks based on WDM technologies forming the core part of Internet infrastructure carrying multi-Tb/s has to be re-considered from both designing, planning, operation and management perspectives so as to support exponential traffic growth in a greater efficiency. Thanks to the convergence of significant advances in optical transmission technologies, and photonic switching, transparent (all-optical) architecture has come into practice, paving the way for eliminating the over-utilization of costly optical-electrical-optical (O-E-O) interfaces and hence, yielding remarkable savings of cost and energy consumption compared to opaque architecture. Traditional designs for transparent optical networks based on single-objective optimization model aiming at optimizing solely a single performance metric appears to be insufficient to capture the nuances of practical designs while conventional multi-objective approach tends to reach (non-) optimal solutions. Different from existing works, we present a new framework for multi-objective WDM network designs capturing several goals on one hand and on the other hand, achieving optimal solutions. Moreover, our proposal exploits the characteristics of each constituent objectives to lay the foundation for setting up weight coefficient so that the order of optimization is guaranteed. Equally important, our proposal is pragmatic in the sense that the complexity of the optimization model remains the same as the single-objective model while the quality of solution has been greatly improved. We have extensively tested realistic optical core networks topologies, that is, COST239 and NSFNET, with various network traffic conditions and it turns out that our design brings about a saving of wavelength link usage up to roughly $28\%$ in the most favorable cases while $14\%$ is expected for the least favorable cases.

\keywords{routing and wavelength assignment \and integer linear programming \and dedicated protection \and optical networks \and fiber optics communication}
\end{abstract}

\section{Introduction}
\label{intro}
The era of Artificial Intelligence and Internet of Things have started to penetrate into the daily lives in an unprecedented way. On one hand, more and more devices have been connected to Internet, paving the way for digitization and then datafication of (almost) everything. On the other hand, the proliferation of bandwidth-demanding services driven by big data analysis, cloud computing and generally digital future have been accelerating in a rapid manner. A case in point for illustration is how data-intensive the operation of a self-driving car is. It is estimated that such an autonomous car generates a huge amounts of data, roughly 5Tb per hour and it is equivalent of a supercomputer in terms of generating and transmitting a mind-boggling amount of data. Such two driving forces have been push-pulling the scale of challenge posed by global consumption of data and the explosive growth of Internet traffic. As recently announced in Cisco Annual Internet Report, the number of devices connected to IP networks will be as more as three times than the global population by 2023, representing a growth of $10\%$ CAGR. As a consequence of such growth, Internet traffic will be increasing three-fold accordingly, reaching a CAGR of $25\%$ and it means that Global Internet traffic in 2020 will be equivalent to 95x the volume of the entire Global Internet in 2005 \cite{Cisco17}. \\

Fiber-optic networks exploiting wavelength division multiplexing (WDM) technologies have long been constituting the core part of Internet infrastructure, supporting the exchange of Internet traffic from a point to any point in the planet thanks to the immense transmission capacity in the order of Tb/s and beyond \cite{20years}. The exponential growth of Internet traffic guided by the rise and growing penetration of bandwidth-intensive services such as virtual realities and AI-enabled applications indeed are posing critical challenges for optical transport networks and hence necessitating for a re-consideration from both designing, planning and management perspectives to achieve greater capital and operational efficiency. From the architectural standpoint, core optical networks have been evolving from the opaque mode to translucent and eventually fully transparent operation. Traditionally, the opaque nodes had been widely implemented for core networks since transmission technologies had been limited to short distances and photonic switching technologies had not been exploited. In this kind of networks, the optical signal are terminated and regenerated at every node by making use of the optical-electrical-optical (O-E-O) conversion. While operating so brings advantage of eliminating cascading physical impairments and allowing multi-vendor interoperability, its critical downsize of over-utilizing costly O-E-O conversion makes it less scalable as the traffic increases. In an attempt to reduce the proliferation of expensive O-E-O usage, translucent networks have been exploited to replace the fully opaque architecture. In this hybrid operation mode, the optical signal remains in optical domain as far as possible before its regeneration (i.e., the regeneration could be either in optical domain or via O-E-O interface). Alternatively, the regeneration could also be employed to ease the wavelength continuity constraints. Noticed that the signal can be regenerated several times in the network before it reaches the destination and hence, could be used different wavelengths at each segment. Thanks to the enormous advancements in optical components, transmission technologies and photonic switching, the vision of all-optical core networks have been experimentally and practically realized, bringing in significant savings of cost, footprint, and power by eliminating unnecessary O-E-O regeneration. Indeed, fully transparent optical core networks nowadays have been widely deployed by network operators and  efficient designing of such transparent networks therefore has been of ever-growing importance as the spectrum resources has to be optimized while maintaining the efficiency of operations. Different from the opaque and translucent networks, operating transparent networks require carefully designed algorithms so that the full benefits of technology could be attained and in this perspective, the essential problem to be solved impacting to all phases of network operations from planning, designing to management is the routing and wavelength assignment (RWA) issue. Basically, RWA involves the finding of a route and assign certain wavelength for a demand so as to optimize a number of performance metrics such as spectrum efficiency, cost or power consumption while meeting technological constraints typically including wavelength continuity and wavelength uniqueness \cite{all-optical, efficient} \\

Being a critical issue for the operation of transparent WDM networks, RWA has been extensively addressed in the literature \cite{rwareview, rwa-protection, rwa-17}. Generally, RWA problem could be formulated in the form of a combinatorial optimization model which could be solved then by either exact approaches to obtain the optimal solutions or heuristic approaches to acquire acceptable and/or near-optimal outcomes \cite{rwa-17}. On one hand, exact approaches based on solving the mixed integer linear programming formulation has the merit of optimality (i.e., higher quality solutions) which could be translated to greater capital expenditure and operation expenditure efficiency. However, the key drawback of such approaches lies in the NP-hard nature of RWA problems and thus, for large-scale networks and/or for certain traffic instances, the solution may be difficult to reach in a practical time frame. On the other hand, effective heuristic algorithms has been sought as an alternative technique aiming at approximated solutions in a rapid manner \cite{mrwa1}. Meta-heuristics extends the quality of general heuristic algorithms by adding specific features of each RWA variants into the algorithms and thus, tailoring the solution to the problem to achieve better performance \cite{mrwa2, mrwa3}. In essence, heuristic-based approaches trade solution qualities for running time in contrast to the exact ones. RWA problems could have different variants depending on the context and the goal of the studies. In the context of survivable network designs, the authors in \cite{1+1rsa, prwa} developed efficient routing strategy to be tolerant to fault issues. Further to the protection scenarios, network coding has been added as a new layer of design and this results in a different variant of RWA, called routing, wavelength and network coding assignment (RWNCA) \cite{hai_comletter, hai_springer, hai_oft, hai_comcom2}. Doing so helps to boost the spectral efficiency of optical core networks by reducing the redundant capacity demanded by protection requirement and the respective mathematical formulation for network design has been introduced in \cite{hai_access, hai_rtuwo}. With respect to energy efficiency, RWA has been re-purposed to make it energy-aware \cite{erwa1, erwa2}. In the paradigm of elastic transmission technnologies, RWA problems are transformed into routing and spectrum assignment and the works in \cite{hai_iet, hai_wiley, hai_csndsp, hai_ps1} proposed new algorithms for solving such related RWA problem in elastic optical networks. A significant part of existing works on RWA have nevertheless been focusing on optimizing a single metric and such conventional metric can be either the number of wavelength\textemdash for the un-bounded scenario to support a given set of traffic demands or alternatively, the throughput in the case of capacity-constrained scenario and hence, the target is to support as many traffic as possible. \\

Planning a network indeed involves a number of metrics, sometimes, conflicting ones and/or with different priorities. Therefore, a multi-objective optimization framework has to be developed to capture the real need of network planning process. In addressing this issue, existing works in the literature have been developed into two directions. The first one is rather straightforward by combining several objectives into a single integrated one with weight coefficients. While this direction appears to be intuitive, existing works in the literature have remained obscured in providing insights on the impact of weight coefficients to the significance of constituents. Besides, the selection of weight coefficients have been largely based on the so-called rule of thumb and hence, difficult to justify. As an alternative approach, the second direction focuses on the application of general multi-objective framework based on meta-heuristic models including swarm intelligence, aiming at obtaining approximate multi-objective solutions (i.e., approximated Pareto front) \cite{mrwa1, mrwa2, mrwa3}. Such approach has nevertheless the essential drawback of missing the particularities of the problem under consideration, in addition to the inherent shortcoming of non-optimal solutions. \\

In mitigating the downsizes of existing approaches for multi-objective RWA design in the literature, this paper goes extra miles with a new integer linear programming model which is capable of optimizing two and more objectives in a strict order. Specifically, the traditional wavelength count metric is optimized as the first priority while wavelength link usage \textemdash being correlated to operational expense\textemdash would be a secondary goal. Such two constituent objectives are combined appropriately into an integrated objective and in guaranteeing the order of optimization, we propose a framework for analyzing and then determining the weight coefficients so that the strict hierarchy between constituent goals are ensured. To this end, the remaining of the paper is organized as follows. In Sect. 2, we present an illustrative example highlighting the key difference of single-objective approach and our multi-objective proposal. Next, in Sect. 3, the new optimization model capturing both wavelength count and wavelength link usage is introduced. We also introduce an extensive analysis of the impact of weight coefficient to the order and priority of constituent objectives. In verifying the efficacy of our proposal, Sect. 4 is dedicated to comprehensive numerical simulations on realistic topologies, COST239 and NSFNET, in different traffic conditions and protection requirements. Finally, Sect. 5 concludes the paper.

\section{Multi-objective transparent WDM network design}
Assuming that there are two traffic requests, from node A to node D and from node E to node D. One such optimal provisioning\textemdash Design A\textemdash for those requests in terms of \textit{wavelength link} is shown in Fig. It is straightforward to verify that the shortest route has been chosen to serve both requests and this design consumes three wavelength links which is an optimal outcome from a conventional single-objective design. On the other hand, such design makes use of two distinctive wavelengths as the routing of connection A-D and E-D share a common link E-D and therefore their wavelengths must be different, that is, $\lambda_1$ and $\lambda_2$ as in Fig. 1a \\

Consider an alternative single-objective design with respect to \textit{wavelength count}\textemdash Design B, one such optimal design is shown in Fig. 1b. It is shown that only one wavelength is needed to accommodate both requests and this is the optimal result obtained from solving the optimization model of such design. However, with respect to \textit{wavelength link} usage, Design B incurs five wavelength links which is roughly $70\%$ higher than the Design A presented in Fig. 1a \\

In improving the Design B in Fig. 1b, one can refer to the Design C shown in Fig. 1c. This Design is a bi-objective model aiming at optimizing the wavelength count first and then the wavelength link usage. It is observed that for Design C, only one wavelength is needed while the wavelength link usage is also minimized, that is, \textit{four} units compared to the (non-optimized) \textit{five} as shown in Design B. Design C therefore on one hand achieves optimal wavelength count as Design B and on the other hand, minimize the wavelength link usage as the secondary objective. The comparison of three Designs is summarized in Tab. 1\\ 

The problem emerged from above analysis is that traditional single-objective designs aim at optimizing a single performance metric and hence possibly incur certain penalties with respect to other objectives. Specifically, it is of particular interest to have a Design that is capable of minimizing multi-objectives in a strict order, that is, the first objective is minimized and then the second objective is optimized among a pool of optimal solutions with respect to the first objective. The same reasoning is applied for general cases of more than two objectives. We aim to propose a general framework for obtaining such pragmatic Designs in this paper.

\begin{figure}[ht]
	\begin{subfigure}{.5\textwidth}
		\centering
		\includegraphics[width=\linewidth]{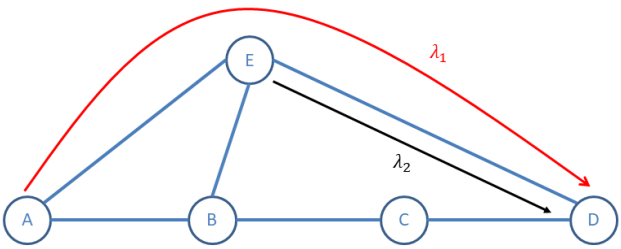}  
		\caption{An optimal single objective design with respect to wavelength link}
		\label{fig:sub-first}
	\end{subfigure}
	\begin{subfigure}{.5\textwidth}
		\centering
		\includegraphics[width=\linewidth]{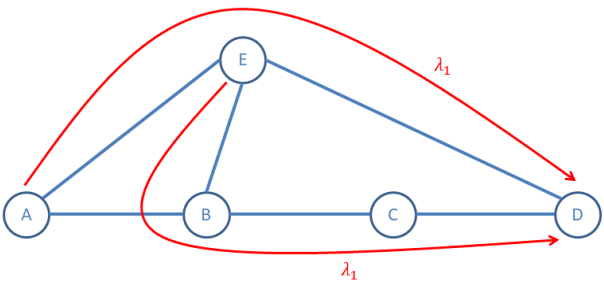}  
		\caption{An optimal single objective design with respect to wavelength count}
		\label{fig:sub-first}
	\end{subfigure}
	\begin{subfigure}{.5\textwidth}
	\centering
	\includegraphics[width=\linewidth]{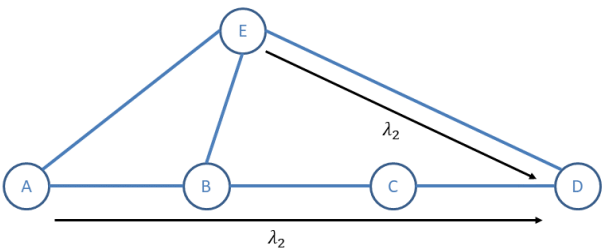}  
	\caption{An optimal bi-objective design with respect to both wavelength count and wavelength link in an ordered fashion}
	\label{fig:sub-first}
	\end{subfigure}
	\caption{An illustrative comparison of different Designs}
	\label{fig:fig}
\end{figure}

\begin{table}[!h]
	\caption{Performance Comparison of Designs}
	\label{tab: Illustration}
	\begin{tabular}{ccc}
		\hline
		Design & \multicolumn{2}{c}{Performance Metric}\\
		\cline{2-3}
		& Wavelength Count & Wavelength Link Usage \\
		\hline
		Design 1 & 2 & 3 \\
		Design 2 & 1 & 4  \\
		Design 3  & 1 & 3 \\
		\hline
	\end{tabular}
\end{table}

\section{Joint Wavelength Count and Wavelength Link Usage Optimization}
\label{sec:1}
From the mathematical perspectives, optical networks design and planning problems belong to the broad class of optimization models. In particular, the planning process of a transport network focuses on how each traffic demand is provisioned so as to minimize/maximize the network performance (e.g., spectrum efficiency). Accommodating a traffic demand then involves the selection of links over which the demand is routed, the backup route for that demand\textemdash if protection service is provided\textemdash in cases of link failures on the working route, the assignment of wavelength for both working and protection signals. When it comes to the design and planning aspects for optical core networks, the timescales for consideration is long-term as the traffic is assumed to be relatively static and this task has to be performed before the network deployment process. In this context, there is generally a large set of demands to be handled simultaneously and the goal is to find optimal strategy to do so. \\

In this part, we present an efficient optimization model for accommodating traffic demands arisen in the long-term planning of optical core networks. Inputs to the formulation are set of nodes and links composing the physical optical network, set of traffic requests and the task is to determine routing for each demand including the working and protection route, the wavelength for both working and protection signals so as to minimize the network resources measured by the number of used wavelength and the wavelength link usage. The fully transparent network architecture is adopted and thus, the wavelength continuity constraint has to be enforced. The network design is formulated in the form of a mixed integer linear programming model composing of an integrated objective function, that is, wavelength count and wavelength link consumption. We present a framework for analyzing and determining the weight coefficients so that the order of optimization goals are ensured. \\
 
\noindent{Given Information:}
\begin{itemize}
	\item $G(V,E)$: Physical fiber-networks topology composing of $|V|$ nodes and $|E|$ fiber links. The starting node and ending node of a link $e \in E$ are denoted as $s(e)$ and $r(e)$, respectively

	\item $D$: Set of all traffic requests. The source node and receiving node of a demand $d\in D$ are denoted as $s(d)$ and $r(d)$ respectively; It is assumed that all demands request \textit{one wavelength capacity}

	\item $C$: Set of all wavelength channels on each fiber link. The fiber link capacity measured by number of wavelength channels is therefore $|C|$
	
	\item $\alpha_1$, $\alpha_2$: Real numbers representing weight coefficients to specify the relative significance of constituent objective 
	
	\item $q_d$: Binary constant specifying the protection service for demand $d$, $q_d=1$ for dedicated protection of demand $d$ while $q_d=0$ implies no protection requirement 
\end{itemize}

\noindent{Variables:}
\begin{itemize}
	\item $x^{e, c}_{d} \in \{0,1\} $: equals to 1 if link $e$ and wavelength channel $c$ is used for working path of demand $d$, 0 otherwise.
	\item $y^{e, c}_{d} \in \{0,1\} $: equals to 1 if link $e$ and wavelength channel $c$ is used for protection path of demand $d$, 0 otherwise.
	\item $\theta ^{c}_{d} \in \{0,1\} $: equals to 1 if the wavelength channel $c$ is used to carry demand $d$, 0 otherwise.
	\item $\gamma_{e, c} \in \{0,1\}$: equals to 1 if the wavelength channel $c$ is used on link $e$, 0 otherwise
	\item $\delta_{c} \in \{0,1\}$: equals to 1 if the wavelength channel $c$ is used in any fiber link of the network, 0 otherwise \\
\end{itemize}

\noindent{Objective: \textit{Minimize the following integrated objective function}
\begin{equation} \label{eq:obj}
 \; \alpha_1 \times \sum_{c \in C} \delta_{c} + \alpha_2 \times \sum_{c \in C} \sum_{e \in E}\gamma_{e, c}
\end{equation}

\noindent{Subject to the following constraints:}
\begin{equation}\label{eq:c1}
\sum_{c \in C} {\theta_d^c} = 1 \; \; \forall d \in D 
\end{equation}

\begin{equation} \label{eq:c2}
\begin{split}
\sum_{e \in {E}: v\equiv s(e)} {x^{e, c}_{d} }-\sum_{e \in {E}: v \equiv r(e)} {x^{e, c}_{d} }= 	
\begin{cases} 
\theta^{c}_{d} &\mbox{if } v \equiv s(d) \\ 
-\theta^{c}_{d}& \mbox{if } v \equiv r(d)\\
$0$ & otherwise \\
\end{cases}   \\  \qquad \qquad \forall v \in V, \forall d \in D, \forall c \in C \hfill
\end{split}
\end{equation}

\begin{equation} \label{eq:c2p}
\begin{split}
\sum_{e \in {E}: v\equiv s(e)} {y^{e, c}_{d} }-\sum_{e \in {E}: v \equiv r(e)} {y^{e, c}_{d} }= 	
\begin{cases} 
\theta^{c}_{d} &\mbox{if } v \equiv s(d) \\ 
-\theta^{c}_{d}& \mbox{if } v \equiv r(d)\\
$0$ & otherwise \\
\end{cases}   \\  \qquad \qquad \forall v \in V, \forall d \in D: q_d = 1, \forall c \in C \hfill
\end{split}
\end{equation}

\begin{align} \label{eq:c4} {
	\sum_{d \in D} x^{e, c}_{d} + \sum_{d \in D} y^{e, c}_{d}  = \gamma_{e,c} \qquad \forall e \in E, \forall c \in C
}
\end{align}

\begin{align} \label{eq:c5} {
	\sum_{c \in C} \gamma_{e, c}  \leq |C| \delta_{c}  \qquad \forall e \in E
}
\end{align}


The integrated objective function in Eq. \ref{eq:obj} consists of two constituents being weighted by coefficients $\alpha_1$ and $\alpha_2$ respectively. The first objective $\delta_c$ is a conventional performance indicator, that is, the number of wavelength channels needed to support traffic demands while the second objective $\gamma_{e, c}$ measures the used wavelength link which directly corresponds to the operational expenses of network when it is in-operation. Eq. \ref{eq:c1} is a constraint to ensure that all demands must be provisioned by having a proper wavelength channel. Eq. \ref{eq:c2} and Eq. \ref{eq:c2p} are flow conservation constraints for working and protection flow respectively. Constraint expressed by Eq. \ref{eq:c4} enforces the uniqueness of each wavelength channel, that is, each wavelength channel is used by at most one lightpath\textemdash either working or protection. The final constraint in Eq. \ref{eq:c5} defines the use of a wavelength channel in a network, that is, a wavelength channel is counted as used when it is assigned for a lightpath in any link of the network. 	

\subsection{Determining Weight Coefficients}

As can be seen in Eq. \ref{eq:obj}, weight coefficients $\alpha_1$ and $\alpha_2$ determine the relative significance of constituent objectives and it would be our interest to find a suitable range of $\alpha_1$ and $\alpha_2$ so as the first objective\textemdash minimizing the wavelength channel count\textemdash is performed with a strict priority and then the second objective\textemdash wavelength link usage\textemdash is optimized. \\

In order to provide such hierarchy of optimization, we exploit the specific characteristic of each constituent objective. In fact, it is known that the first objective is lower-bounded and upper-bounded as $0 <= \sum_{c \in C}\delta_c <= |C|$. Similarly, the second objective also has a lower bound and upper-bound as $0 <= \sum_{c \in C} \sum_{e \in E} \gamma_{c, e} <= |C||E|$. Next the minimum step of variation for each objective is also known, that is, \textit{one}. Let us consider the case that $\alpha_1 > |E||C| \times \alpha_2$. \\

We assume that there is a minimum increase from the first objective, that is, from $\delta_c$ to $\delta_c + 1$ and there is an arbitrary change $\triangle$ in the second objective. The net impact on the integrated objective function is therefore $N = \alpha_1 \times (\delta_c + 1 - \delta_c) + \alpha_2 \times \triangle$. Conditioned on $-|E||C| <= \triangle <= |E||C|$ and $\alpha_1 > |E||C| \times \alpha_2$, it is easy to verify that $N$ is always greater than zero. This observation reveals the fact that variation of the integrated objective function is in the same direction with the first constituent objective despite arbitrary deviation of the second objective. More specifically, under the case that $\alpha_1 > |E||C| \times \alpha_2$, there is a strict order of optimization, that is, the first objective is optimized first and then the second objective is taken into consideration.  

\section{Numerical Evaluations}

This section is dedicated to provide a comprehensive evaluation of our design proposal in realistic network topologies, COST239 and NSFNET, in various traffic conditions. The key characteristic of these networks are reported in Tab. 2. In order to evaluate the efficacy of the integrated objective design, it is benchmarked with the traditional single-objective design that is widely used in the literature. In particular, four Designs are under comparison. The first and second Design are the conventional single-objective RWA in the case of no protection and dedicated protection respectively (i.e., rwa\_wc and rwa\_wc\_p). The third and fourth design are the (multi-) bi-objective RWA (i.e., rwa\_intwc and rwa\_intwc\_p) aiming at minimizing the wavelength count first and wavelength link usage as a secondary goal. Description of these four Designs together with their objective functions are shown in Tab. 3. To represent different traffic conditions, we consider three traffic loads, that is, low, medium and high whose difference is the number of traffic demands. The high load is characterized by the fact that all the node pairs in the network have traffic requests while for the low and medium ones, $\approx 30\%$ and $\approx 70\%$ of all node pairs have traffic respectively. For each type of traffic load, 10 instances of traffic matrices are generated for simulations. In solving the optimization model based on mixed integer linear programming of four designs, we use CPLEX solver version 12.7 with academic license. All the results are optimally obtained and in most demanding cases, the running time is less than 1 hour. \\

Table 4 presents a comprehensive simulation results comparing the traditional single-objective Designs and our proposal of multi-objective Designs in two scenarios, that is, with dedicated protection and without protection. The performance metric for comparison are wavelength count and wavelength link usage. Note that the results are averaged over 10 traffic instances.

On a first observation, it is noted that the integrated objective Designs achieve the same quality of solution compared to the traditional single-objective Designs with respect to wavelength count metric and this is ensured consistently over different load conditions and for both two networks (i.e., between rwa\_pc and rwa\_intwc on wavelength count, between rwa\_wc\_p and rwa\_intwc\_p on wavelength count). This is the experimental confirmation from our theoretical analysis in Sect. 3. Note that such same quality of solution could be achieved only by proper set-up of the weight coefficients and we have demonstrated a framework for doing so.  \\

As can be seen in Tab. 4, adding protection requirement in provisioning traffic requests have caused redundant capacity including extra spectrum capacity and/or hardware equipments. Interestingly, such impact is different depending on traffic load conditions. For COST239 network, in the low-load condition, an increase of $85\%$ wavelength count has to be incurred while in medium and high-load, the impact is more severe, being $130\%$ and $171\%$ respectively. For NSFNET network which is sparser than COST239 network, the cost of having protection is very high and it is relatively consistent over three different traffic loads. Specifically, with respect to wavelength count, adding protection demands an increase of more than $200\%$ wavelengths. \\

Next, the important observation involves how the proposal of integrated objective Designs is more efficient than the traditional single objective approach. This can be verified by focusing on the comparison of wavelength link usage. As has been clearly shown from results in Tab. 4, the integrated objective approach consistently outperforms in all traffic load conditions and for both cases of protection or without protection. Specifically, at low-load, the integrated Designs (i.e., rwa\_intwc and rwa\_intwc\_p) yield $14\%$ and $20\%$ savings of wavelength link for COST239 and NSFNET networks respectively. For the case of protection, that saving is more significant, being $19\%$ for COST239 and $28\%$ for NSFNET network. The highest saving is recorded at medium-load condition as our proposal is $20\%$ to $22\%$ more efficient than its counterpart for both cases of protection or without protection and for both COST239 and NSFNET network.   \\

\begin{figure}[!ht]
	\centering
	\includegraphics[width=\linewidth]{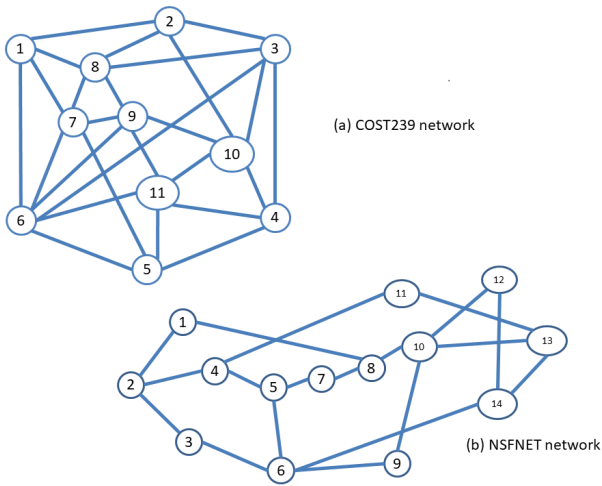}
	\caption{Network topologies for evaluation}
	\label{fig:topology}
\end{figure}

\begin{table}[!ht]
	\caption{Topology Characteristic}
	\label{tab: Topology Info}
	\begin{tabular}{ccc}
		\hline
		Parameters & \multicolumn{2}{c}{Network Topology}\\
		\cline{2-3}
		& COST239 & NSFNET \\
		\hline
		 Nodes & 11 & 14 \\
		 Links & 26x2 & 21x2  \\
		Average degree of nodes & 3 & 3.4 \\
		Min node degree  & 2 & 3 \\
		Max node degree  & 4 & 5 \\
		\hline
	\end{tabular}
\end{table}

\begin{table*}[!ht]
	\caption{Description of Four Designs for Benchmarking}
	\label{tab:result0}
	\centering
	\begin{tabular}{ccc}
		\hline
		Design Name & Objective Function & Description \\
		\hline		
		rwa\_wc & $\sum_{c \in C} \delta_{c}$ & Wavelength count is a sole objective for minimization \\
		& & There is no protection in provisioning a demand \\ 
		\hline
		rwa\_wc\_p & $\sum_{c \in C} \delta_{c}$ & Wavelength count is a sole objective for minimization \\
		& & There is a dedicated protection in provisioning a demand \\
		\hline
		rwa\_intwc &  $\sum_{c \in C} \delta_{c} + \frac{1}{1+|C||E|} \sum_{c \in C} \sum_{e \in E}\gamma_{e, c}$  & Minimize wavelength count first and then wavelength link usage\\
		& & There is no protection in provisioning a demand \\ 
		\hline
		rwa\_intwc\_p &  $\sum_{c \in C} \delta_{c} + \frac{1}{1+|C||E|} \sum_{c \in C} \sum_{e \in E}\gamma_{e, c}$  & Minimize wavelength count first and then wavelength link usage \\
		& & There is dedicated protection in provisioning a demand \\ 
		\hline
	\end{tabular}  
\end{table*}

\begin{table*}[!ht]
	\caption{Performance Comparison of Four Designs on COST239 and NSFNET topology}
	\label{tab:result1}
	\centering
	\begin{tabular}{ccccc}
		\hline
		Network Topology & Traffic Condition & Design Name &  \multicolumn{2}{c}{Performance Metrics} \\
		\cline{4-5}
		&         &          &  Wavelength Count & Wavelength Link Usage \\         
		
		\hline		
		COST239 & Low-load & rwa\_wc & 1.4  & 33.2\\
		
			& & rwa\_wc\_p & 2.6 & 85.3 \\
		& & rwa\_intwc & 1.4 & 28.7 \\
		& & rwa\_intwc\_p & 2.6 & 69.1 \\
		
		\cline{2-5}
		& Medium-load & rwa\_wc & 2.0  & 73.5\\
		
			& & rwa\_wc\_p & 4.6 & 169.3 \\
		& & rwa\_intwc & 2.0 & 57.4 \\
		& & rwa\_intwc\_p & 4.6 & 136.1 \\
	
		\cline{2-5}
		& High-load  & rwa\_wc & 2.8  & 109.3\\
		
			& & rwa\_wc\_p & 7.6 & 278.3 \\
		& & rwa\_intwc & 2.8 & 87.7 \\
		& & rwa\_intwc\_p & 7.6 & 218.0\\
		
		\hline
		NSFNET & Low-load & rwa\_wc & 3.4 & 83.7\\
		
			& & rwa\_wc\_p & 10.5 & 246.5 \\
		& & rwa\_intwc & 3.4 & 67.0 \\
		& & rwa\_intwc\_p & 10.5 & 178.7\\
		
		\cline{2-5}
		& Medium-load & rwa\_wc & 5.6 & 160.9\\
	
			& & rwa\_wc\_p & 19.6 & 461.8 \\
		& & rwa\_intwc & 5.6 & 132.4 \\
		& & rwa\_intwc\_p & 19.6 & 377.9 \\
		
		\cline{2-5}
		& High-load & rwa\_wc & 7.3  & 230.0\\
		
			& & rwa\_wc\_p & 24.7 & 676.4 \\
		& & rwa\_intwc & 7.3 & 198.6 \\
		& & rwa\_intwc\_p & 24.7 & 586.3 \\
		
		\hline
	\end{tabular}  
\end{table*}

\section{Conclusions}

This paper set out to investigate a different approach for solving the multi-objective RWA problem arisen in the designing and planning of transparent WDM optical networks. Inspired by the observation that, in practice, in addition to the key design metric, other performance metrics should also be included for consideration and hence, being optimized in the optimization model with different priorities. In doing so, we have presented a new framework for combining multiple objectives into a single integrated objective function with appropriate weight coefficients. The priority of each constituent objective is determined by its weights and we have provided a comprehensive analysis on the impact of weight coefficients to their priorities. Extensive numerical simulations have been carried out on realistic network topologies including COST239 and NSFNET, with various traffic conditions and protection requirements and it has been highlighted that our proposal could yield a saving of up to $28\%$ wavelength link usage in most favorable cases while in the worst cases, the saving has remained roughly $14\%$. The contribution of this paper\textemdash the network design formulation\textemdash could therefore provide insights for network operators in redesigning their networks to achieve greater capital and operational efficiency. \\

In extending this work, one promising direction would be to translate our proposal to  more practical performance indicators including cost and energy consumption \textemdash which of course depending on the technological and network settings, in addition to including more than two objectives in the integrated objective function. Further research might also explore the multi-objective designs in the context of advanced technological and architectural frameworks including photonic network coding and/or space division multiplexing-elastic optical networks. 

\section*{Conflict of interest}
The authors declare that they have no conflict of interest.


\bibliographystyle{spmpsci_modified}      
\bibliography{ref}   

\end{document}